\documentclass[sigconf,review=false]{acmart}

\usepackage{booktabs} 
\usepackage{subcaption}
\usepackage{cleveref}
\newcommand{\etal}{\textit{et al. }}
\newcommand{\projname}{User Activity Jaccard Similarity }

\setcopyright{none}

\acmDOI{ }

\acmISBN{ }

\acmConference[]{}{} 
\acmYear{ }
\copyrightyear{ }

\acmPrice{ }

\begin{document}
\title[A quantitative, cross-platform analysis of the Italian Referendum]{Public discourse and news consumption on online social media: A quantitative, cross-platform analysis of the Italian Referendum}

\author{Michela Del Vicario}
\orcid{}
\affiliation{
  \institution{IMT School for Advanced Studies}
  \city{Lucca}
  \country{Italy}
}
\email{michela.delvicario@imtlucca.it}

\author{Sabrina Gaito}
\affiliation{
  \institution{University of Milan}
  \department{Dept. of Computer Science}
  \city{Milan} 
  \state{Italy} 
}
\email{sabrina.gaito@unimi.it}	

\author{Walter Quattrociocchi}
\affiliation{
  \institution{IMT School for Advanced Studies}
  \city{Lucca}
  \country{Italy}
}
\email{walter.quattrociocchi@imtlucca.it}

\author{Matteo Zignani}
\affiliation{
	\institution{University of Milan}
	\department{Dept. of Computer Science}
	\city{Milan} 
	\state{Italy}
}
\email{matteo.zignani@unimi.it}

\author{Fabiana Zollo}
\affiliation{%
	\institution{Ca' Foscari University of Venice}
	\department{DAIS}
	\state{Venice}
	\country{Italy} 
}
\email{fabiana.zollo@unive.it}

\renewcommand{\shortauthors}{M. Del Vicario \etal}

\begin{abstract}
The rising attention to the spreading of fake news and unsubstantiated rumors on online social media and the pivotal role played by confirmation bias led researchers to investigate different aspects of the phenomenon. Experimental evidence showed that confirmatory information gets accepted even if containing deliberately false claims while dissenting information is mainly ignored or might even increase group polarization. It seems reasonable that, to address misinformation problem properly, we have to understand the main determinants behind content consumption and the emergence of narratives on online social media. In this paper we address such a challenge by focusing on the discussion around the Italian Constitutional Referendum by conducting a quantitative, cross-platform analysis on both Facebook public pages and Twitter accounts. We observe the spontaneous emergence of well-separated communities on both platforms. Such a segregation is completely spontaneous, since no categorization of contents was performed a priori.  By exploring the dynamics behind the discussion, we find that users tend to restrict their attention to a specific set of Facebook pages/Twitter accounts. Finally, taking advantage of automatic topic extraction and sentiment analysis techniques, we are able to identify the most controversial topics inside and across both platforms. We measure the distance between how a certain topic is presented in the posts/tweets and the related emotional response of users. Our results provide interesting insights for the understanding of the evolution of the core narratives behind different echo chambers and for the early detection of massive viral phenomena around false claims.
\end{abstract}

%
%
%

\keywords{information spreading, news consumption, cross-platform comparison, online social media}

\maketitle

\section{Introduction}

Social media have radically changed the paradigm of news consumption and their impact on the information spreading and diffusion has been largely addressed  \cite{asur2010predicting,becker2010learning, ju2014will, hermida2010twittering, castillo2014characterizing}. In particular, several studies focused on the prediction of social dynamics \cite{asur2010predicting,becker2010learning}, with a special emphasis on information flows patterns \cite{leskovec2011social} and on the emergence of specific community structures \cite{grabowicz2012social, papadopoulos2012community}. Indeed, social media are  always more involved in both the distribution and consumption of news \cite{olteanu2015comparing}. According to a recent report \cite{newman2015reuters}, approximately 63\% of users access news directly from social media, and such information undergoes the same popularity dynamics as other forms of online contents, such as selfies or kitty pictures. In such a disintermediated environment, where users actively participate to contents production and information is no longer mediated by journalists or experts, communication strategies have changed, both in the way in which messages are framed and are shared across the social networks. The general public drifts into dealing with a huge amount of information, but the quality may be poor. Social media do represent a great tool to inform, engage and mobilize people easily and rapidly. However, it also has the power to misinform, manipulate or control public opinion. Since 2013, the World Economic Forum (WEF) has been listing massive digital misinformation at the core of technological and geopolitical risks to our society, along with terrorism and cyber attacks \cite{howell2013digital}. Indeed, online disintermediation elicits users' tendency to select information that adheres (and reinforces) their pre-existing beliefs, the so-called confirmation bias. In such a way, users tend to form groups of like-minded people where they polarize their opinion, i.e., echo chambers \cite{del2016spreading,quattrociocchi2016echo,bessi2015viral,zollo2015debunking}. Confirmation bias plays a pivotal role in informational cascades. Experimental evidence shows that confirmatory information gets accepted even if containing deliberately false claims \cite{bessi2014science,bessi2015viral}, while dissenting information is mainly ignored \cite{zollo2015debunking}. Moreover, debating with like-minded people has been shown to influence users' emotions negatively and may even increase group polarization \cite{sunstein2002law, zollo2015emotional}. Therefore, current approaches such as debunking efforts or algorithmic-driven solutions based on the reputation of the source seem to be ineffective to contrast misinformation spreading \cite{qazvinian2011rumor}. 

The rising attention to the spreading of fake news and unsubstantiated rumors on online social media led researchers to investigate different aspects of the phenomenon, from the characterization of conversation threads \cite{backstrom2013characterizing}, to the detection of bursty topics on microblogging platforms \cite{diao2012finding}, to the mechanics of diffusion across different topics \cite{romero2011differences}. Misinformation spreading also motivated major corporations such as Google and Facebook to provide solutions to the problem \cite{firstdraftnews}. Users' polarization is a dominating aspect of online discussions \cite{adamic2005political,ugander2012structural}. Furthermore, it has been shown that on Facebook users tend to confine their attention on a limited set of pages, thus determining a sharp community structure among news outlets \cite{schmidt2017anatomy}, and a similar pattern was also observed around the Brexit debate \cite{del2016anatomy}. Therefore, it seems reasonable that, to address misinformation problem properly, we have to understand the main determinants behind content consumption and the emergence of narratives on online social media. In this paper we address such a challenge by focusing on the discussion around the Italian Constitutional Referendum by conducting a quantitative, cross-platform analysis on both Facebook public pages and Twitter accounts. First, we characterize the structural properties of the discussion by modeling users' interactions by means of a bipartite network where nodes are Facebook pages (respectively, Twitter accounts) and connections among pages (respectively, accounts) are the direct result of users' activity. By comparing the results of  different community detection algorithms, we are able to determine the existence of well-separeted communities on both platforms. Notice that such a segregation is completely spontaneous, since we did not perform any categorization of contents a priori. Then, we explore the dynamics behind the discussion by looking at the activity of the users within the most active communities, finding that users tend to restrict their attention to a specific set of pages/accounts.  
Finally, taking advantage of automatic topic extraction and sentiment analysis techniques, we are able to identify the most controversial topics inside and across both platforms. We measure the distance between how a certain topic is presented in the posts/tweets and the related emotional response of users.
Summarizing, the novel contribution of this paper is double:
\begin{enumerate}
\item First, we provide a comparative, quantitative analysis of the way in which news and information get consumed on two very different and popular platforms, Facebook and Twitter; 
\item Second, we provide automatic tools to detect the popularity of narratives online and their related polarization effects.
\end{enumerate}
Such an approach may provide important insights for the understanding of the evolution of the core narratives behind different echo chambers and for the early detection of massive viral phenomena around false claims.

\section{Dataset}
Facebook and Twitter are two of the most popular online social media where people can access and consume news and information. However, their nature is different: Twitter is an information network, while Facebook is still a social network, despite its evolution into a "personalized portal to the online world" \footnote{\url{http://www.slate.com/articles/technology/technology/2016/04/facebook_isn_t_the_social_network_anymore_so_what_is_it.html}}. Such a difference highlights the importance of studying, analyzing, and comparing users' consumption patterns inside and between both platforms.
\subsection{Facebook}
Following the exhaustive list provided by \textit{Accertamenti Diffusione Stampa} (ADS) \cite{ads}, we identified a set of $57$ Italian news sources and their respective Facebook pages. For each page, we downloaded all the posts from July 31st to December 12th, 2016, as well as all the related likes and comments. Then we filtered out posts about the Italian Constitutional Referendum (hold on December, 4th) by keeping those containing at least two words in the set \{\textit{Referendum, Riforma, Costituzionale}\} in their textual content i.e., their description on Facebook or the URL content they pointed to. The exact breakdown of the dataset is provided in Table \ref{tab:basic_stats}. Data collection was carried out using the Facebook Graph API \cite{fb_graph_api}, which is publicly available. For the analysis (according to the specification settings of the API) we only used publicly available data (thus users with privacy restrictions are not included in the dataset). The pages from which we downloaded data are public Facebook entities and can be accessed by anyone. Users' content contributing to such pages is also public unless users' privacy settings specify otherwise, and in that case it is not available to us.

\begin{table}[h!]
	\centering
	\caption{Breakdown of Facebook and Twitter datasets.}
	\label{tab:basic_stats}
	\begin{tabular}{lcc}
		\toprule
		 & Facebook & Twitter\\
		\midrule
		\texttt{Pages/Accounts} & 57 & 50 \\
		\texttt{Posts/Tweets} & 6015 & 5483\\
		\texttt{Likes/Favorites} & 2034037 & 57567 \\
		\texttt{Retweets} & - & 55699\\
		\texttt{Comments} & 719480 & 30079 \\
		\texttt{Users} & 671875 & 29743 \\
		\texttt{Likers/Favoriters} & 568987 & 16438\\
		\texttt{Retweeters} & - & 14189 \\
		\texttt{Commenters} & 200459 & 8099 \\
		\bottomrule
	\end{tabular}
\end{table}

\subsection{Twitter}
Data collection on Twitter followed the same pipeline adopted for Facebook. First, we identified the official accounts of all the news sources reported in \cite{ads}. Then, we gathered all the tweets posted by such accounts from July 31st to December 12th, 2016 through the Twitter Search API. Starting from that set, we selected only tweets pointing to news related to the Referendum debate. Specifically, we considered only the statutes\footnote{Status and tweet are synonyms.} whose URL was present in the Facebook dataset. The resulting set of valid tweets consists of about 5,400 elements. To make the available information comparable to Facebook, from each tweet we collected information about the users who "favored" (left a favorite), retweeted or replied. Specifically, favorites and retweets express an interest in the content as Facebook likes do, while replies are similar to comments. To obtain such information we bypassed the Twitter API, since it does not return favorites or retweets. In particular, we scraped the tweet page and extracted the users who liked or retweeted the status, and the retweet/favorite counts. Although we are able to identify the users acting on a tweet, the information is partially incomplete since the tweet page shows at most 25 users who retweet or favor (retweeters or "favoriters"). However, such a restriction has a limited impact on the set of valid tweets. Indeed, our retweeters' and favoriters' sets capture the entire set of users acting on tweets for about $80\%$ of statuses. As for the replies to a tweet, we were able to collect every user who commented on a tweet reporting a news about the Referendum and the reply tweet she wrote. The tweet page reports the entire discussion about the target tweet, including the replies to replies and so on. Here we limited our collection to the first level replies --i.e., direct replies to the news source tweet-- since the target of the comment is identifiable, i.e. the news linked to the tweet\footnote{For deeper levels the target could be the parent reply or the target tweet.}. 

The breakdown of the Twitter dataset is reported in Table\ref{tab:basic_stats}. The ratio between Facebook and Twitter volumes mirrors the current social media usage in the Italian online media scene. Indeed, news sources and newspapers exploit both media channels to spread their contents, as denoted by the similar number of posts. Nevertheless, activities on the posts are skewed towards Facebook, since in Italy its active users are 4/5-fold those of Twitter \footnote{\url{http://vincos.it/2016/04/01/social-media-in-italia-analisi-dei-flussi-di-utilizzo-del-2015/}}.
Our collection methodology and the high comparability between the two datasets provide a unique opportunity to investigate news consumption inside and between two different and important social media.

\section{Methods}
Interactions between Facebook pages/Twitter accounts and users can be modeled by a bipartite network $G=(P,U,E)$, where $P$ and $U$ are disjoint sets of vertices representing news sources accounts and users, respectively. When a user $u\in U$ interacts with a post/tweet published by the news source $p\in P$, we draw a link $(u,p)\in E$. To each link we assign a weight $w_p^u$ denoting how many times $u$ has interacted with the posts published by the page $p$. Since we distinguish between comments/replies and likes/favorites-retweets, we have different types of weight corresponding to different kinds of interaction.
\\Starting from the bipartite network representation, we are then able to extract groups of news sources which are perceived as similar by social media users.

\subsection{Bipartite Network Projections}
When limiting to a specific topic --i.e., the constitutional Referendum-- a single user interacting with different news sources can be interpreted as a signal of closeness or similarity between the pages, especially when interactions are likes, favorites or retweets. To this aim, we apply three different projections of the bipartite network of pages and users on the pages set, where in turn we assign a different weight to the inter-pages links, assessing diverse similarity scores:
\begin{itemize}
	\item \textbf{Simple Edge Counting}: The weight of each link is given by the number of common users between two pages, in terms of either likes (Facebook) or favorites and retweets (Twitter); 
	\item \textbf{Jaccard Similarity}: Given two pages $a$ and $b$, and their respective sets of users $A$ and $B$, the weight of their link is given by the Jaccard similarity coefficient of the two pages $J_{ab} = |A\cap B|/|A \cup B|$;
	\item \textbf{\projname}: We also consider a modification of the Jaccard Similarity by taking into account the weight of the page-user links. Let $U_{ab}$ be the set of common users between pages $a$ and $b$, whose cardinality is $N_{ab}$, $U$ the set of all users, and $P$ that of all pages, the page relative users weight $w_{ab}$ is equal to:
	$$
	w_{ab} = \frac{1}{N_{ab}}	\sum_{u \in U_{ab}} \left( 1 - \frac{|w_a^u - w_b^u|}{\max\limits_{c\in P}w_c^u} \right) \frac{|w_a^u + w_b^u|}{2\max\limits_{v \in U, c \in P}w_c^v}.
	$$
	The \projname coefficient is given by the product $w_{ab}J_{ab}$.
\end{itemize}

\begin{figure*}[t!]
	\begin{subfigure}{0.9\columnwidth}
		\centering
		\includegraphics[width=\textwidth]{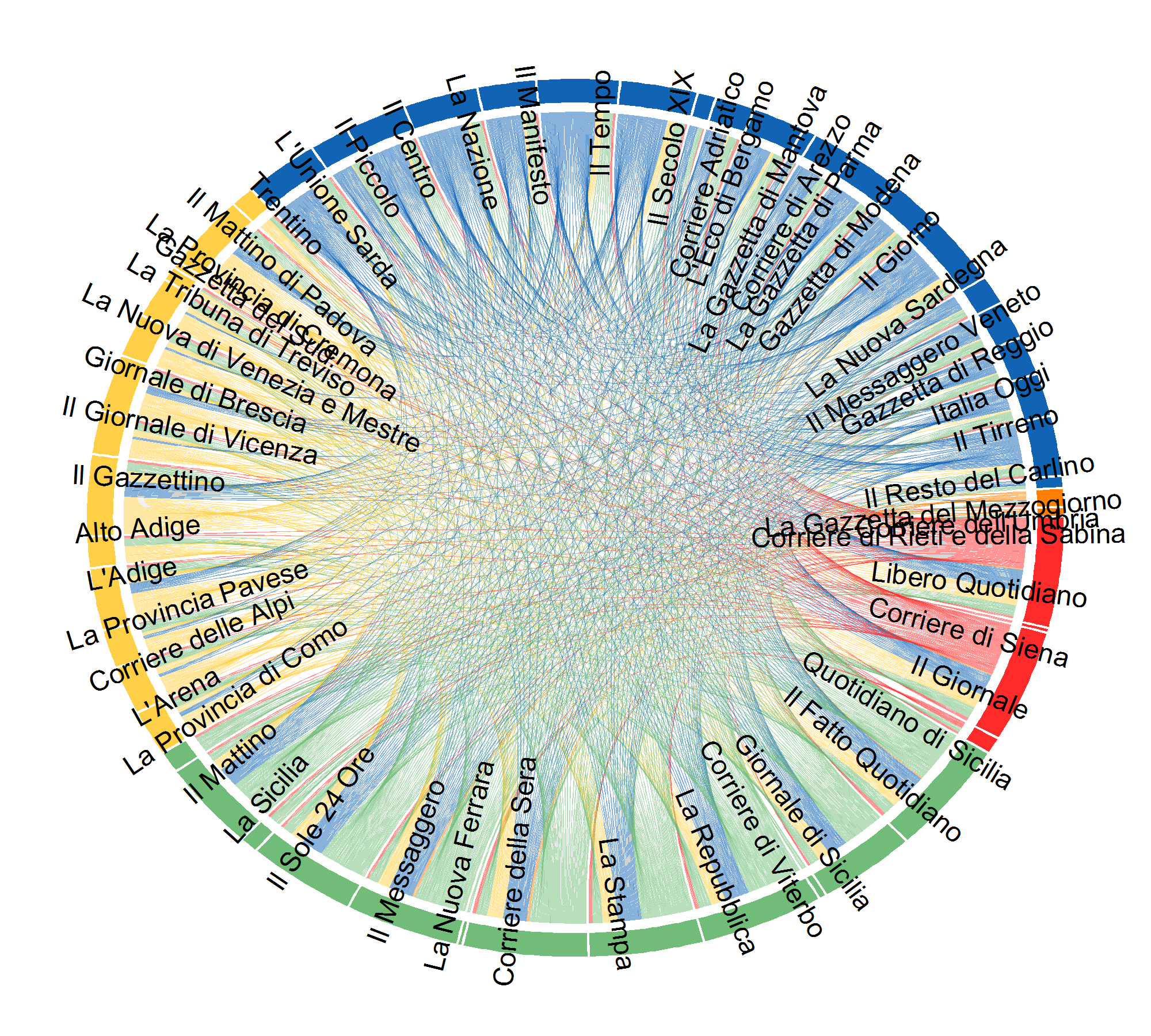}
		\caption{Facebook communities}
		\label{fig:facebok_coms}
	\end{subfigure}
	\begin{subfigure}{0.9\columnwidth}
		\centering
		\includegraphics[width=\textwidth]{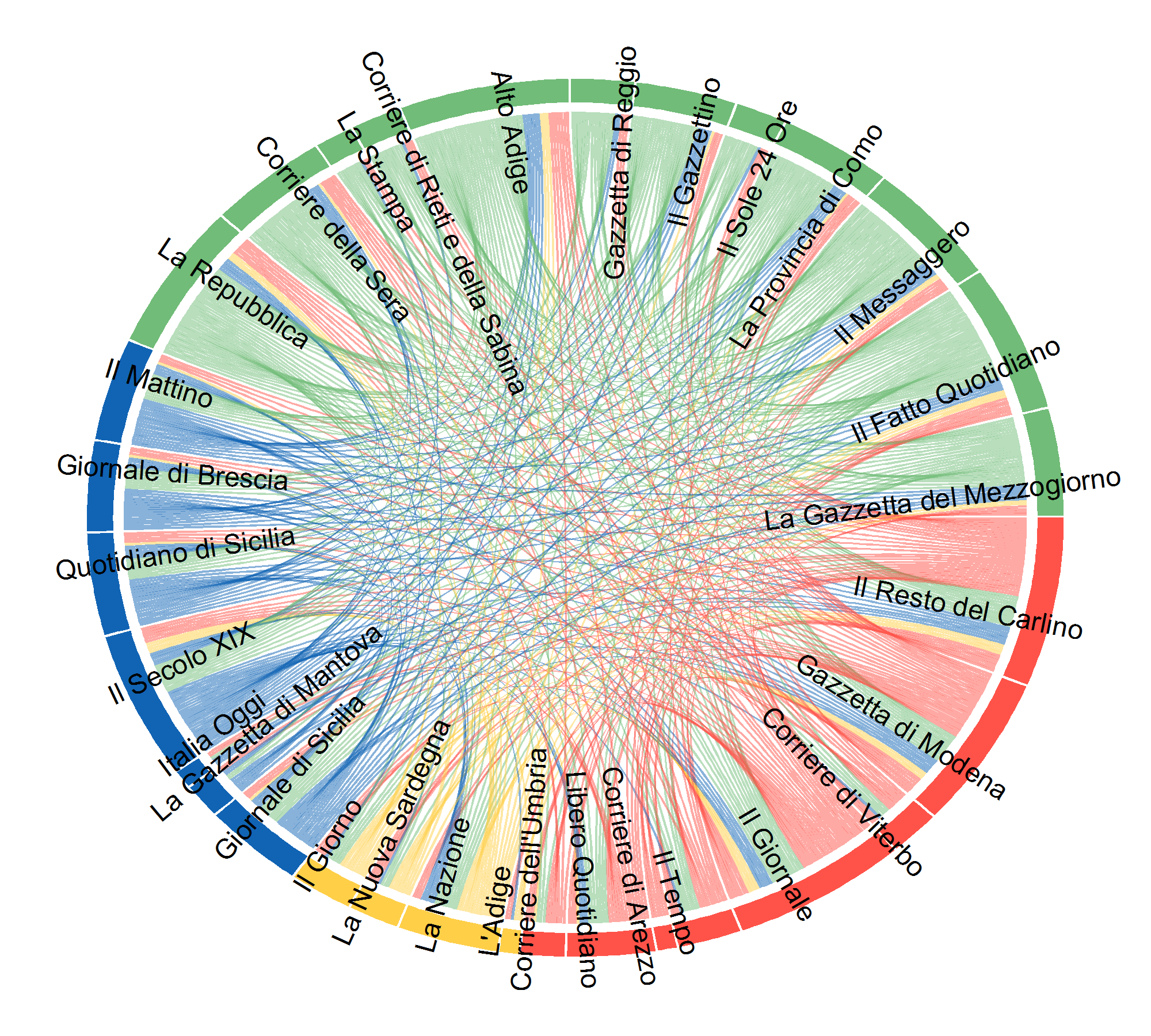}
		\caption{Twitter communities}
		\label{fig:twitter_coms}
	\end{subfigure}
	\caption{a) Community structure for the \projname projection of the pages-users graph on Facebook and b) on Twitter. Colors indicate the membership of Facebook pages or Twitter accounts in the different communities (green for C1, red for C2, blue for C3, yellow for C4, and orange for C5) detected by the \textit{Louvain algorithm}.}
	\label{fig:commmunities}
\end{figure*}

\subsection{Community detection algorithms}
Given the above projections, we are able to identify groups of newspapers which are perceived similar by the social media users. The identification of the groups turns into the well-known community detection problem. For this task, we choose algorithms representative of different approaches on community detection, ranging from optimizing the modularity function, to dynamical processes such as diffusion or random walks.
\begin{itemize}
	\item \textbf{Fast Greedy ~\cite{ClausetPRE2004}:} The algorithm is based on the maximization of the Newman and Girvan's modularity, a quality function which measures the goodness of a graph clustering w.r.t. its random counterpart with the same expected degree sequence. The algorithm starts from a set of isolated nodes and, at each step, continues to add links from the original graph to produce the largest modularity increase.
	\item \textbf{Louvain ~\cite{BlondelJSM2008}:} The Louvain method was introduced by Blondel \etal and is one of the most popular greedy algorithms for \emph{modularity} optimization. The method is very fast and can produce very high quality communities.
	The algorithm is iterative. Each iteration consists of two steps. In the first one, every node is initially set to a new community. Then, for every node $i$ and its neighbors $j$, the algorithm calculates the gain in modularity  moving $i$ from its community to $j$'s community. The node $i$ is then moved to the community with maximum gain. The second step is to group all the nodes in the same community into a macro node.  A new graph, in which macro nodes are linked by an edge if there's an edge between two nodes belonging to the two macro nodes, is built and a new iteration starts.
	\item \textbf{Label propagation ~\cite{RaghavanPRE2007}:} The algorithm provides each node $v \in V$ to determine its community by choosing the most frequent label shared by its neighbors. Initially every node belongs to a different community.  After some iterations, groups of nodes quickly reach a consensus on their label and they begin to contend those nodes that lay between groups.
	\item \textbf{Walk Trap ~\cite{PonsJGAA2006}:} The algorithm is based on the tendency of random walks to be trapped into the densely connected regions of the graph, i.e. communities. Specifically, a random walk of length $t$ induces a distance between two vertices depending on the probability the random walker has to go from a vertex to the other in $t$ steps. The distance can be used in a hierarchical clustering algorithm to obtain a hierarchical community structure.
	\item \textbf{Infomap ~\cite{RosvallEPJ2009}:} Infomap turns the community detection problem into the problem of optimally compressing the information of a random walk taking place on the graph. The algorithm achieves the optimal compression by optimizing the minimum description length of the random walk. 	
\end{itemize}
To compare the partitions returned by the different algorithms, we use standard techniques that compute the similarity between different clustering methods by considering how nodes are assigned by each community detection algorithm \cite{rand1971objective,hubert1985comparing}. Specifically, we adopted the Rand Index \cite{rand1971objective}, based on the relationships between pairs of vertices in both partitions, and the Normalized Mutual Information (NMI) \cite{danon2005comparing}, measuring the statistical dependence between the partitions. In both cases a value close to 1 indicates that the partitions are almost identical, whereas independent partitions lead the scores to 0.

\section{Results}

\begin{figure*}[h!]
	\begin{subfigure}{1\columnwidth}
		\centering
		\includegraphics[width=\textwidth]{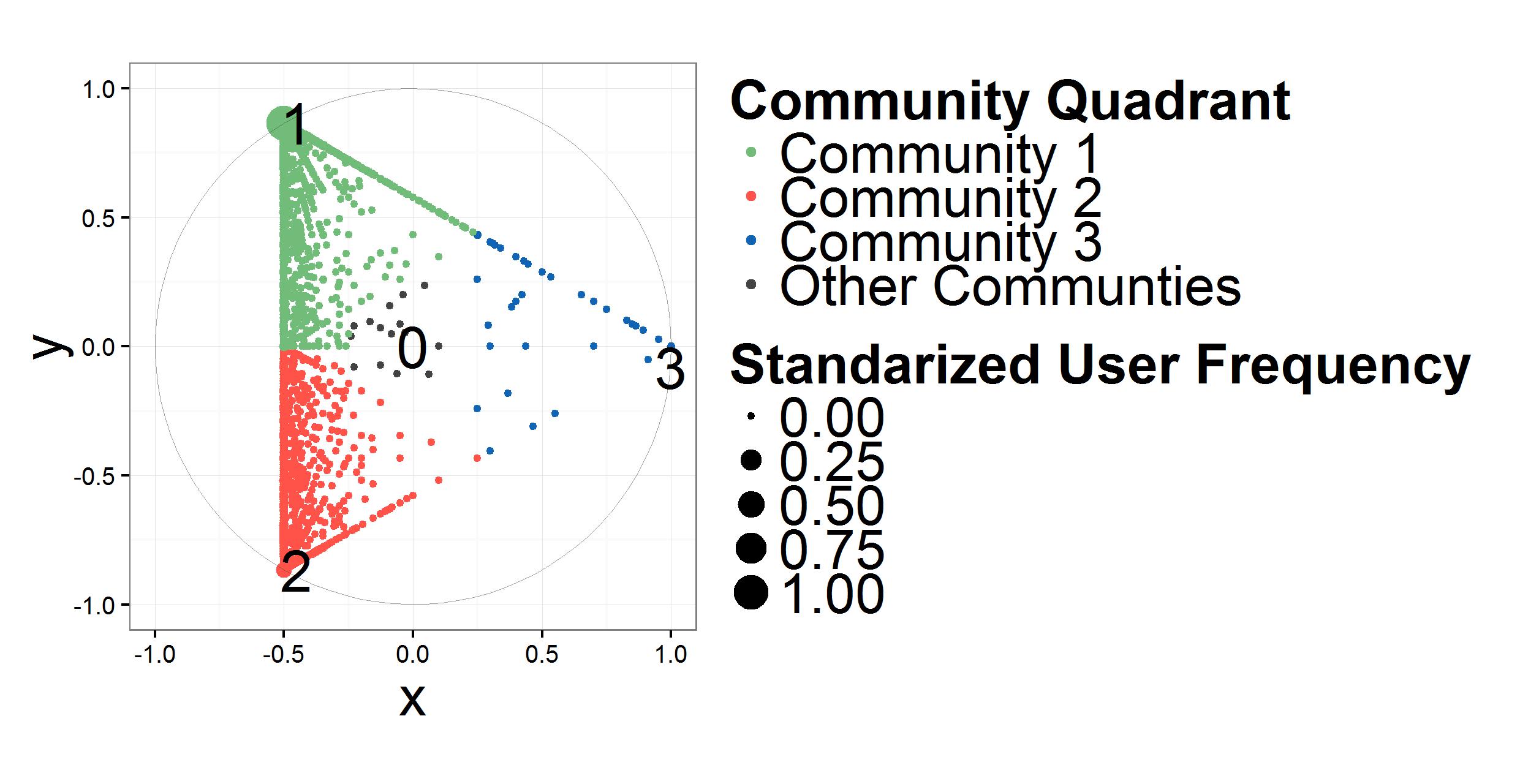}
		\caption{Facebook}
		\label{fig:facebok_polarization}
	\end{subfigure}
	\begin{subfigure}{1\columnwidth}
		\centering
		\includegraphics[width=\textwidth]{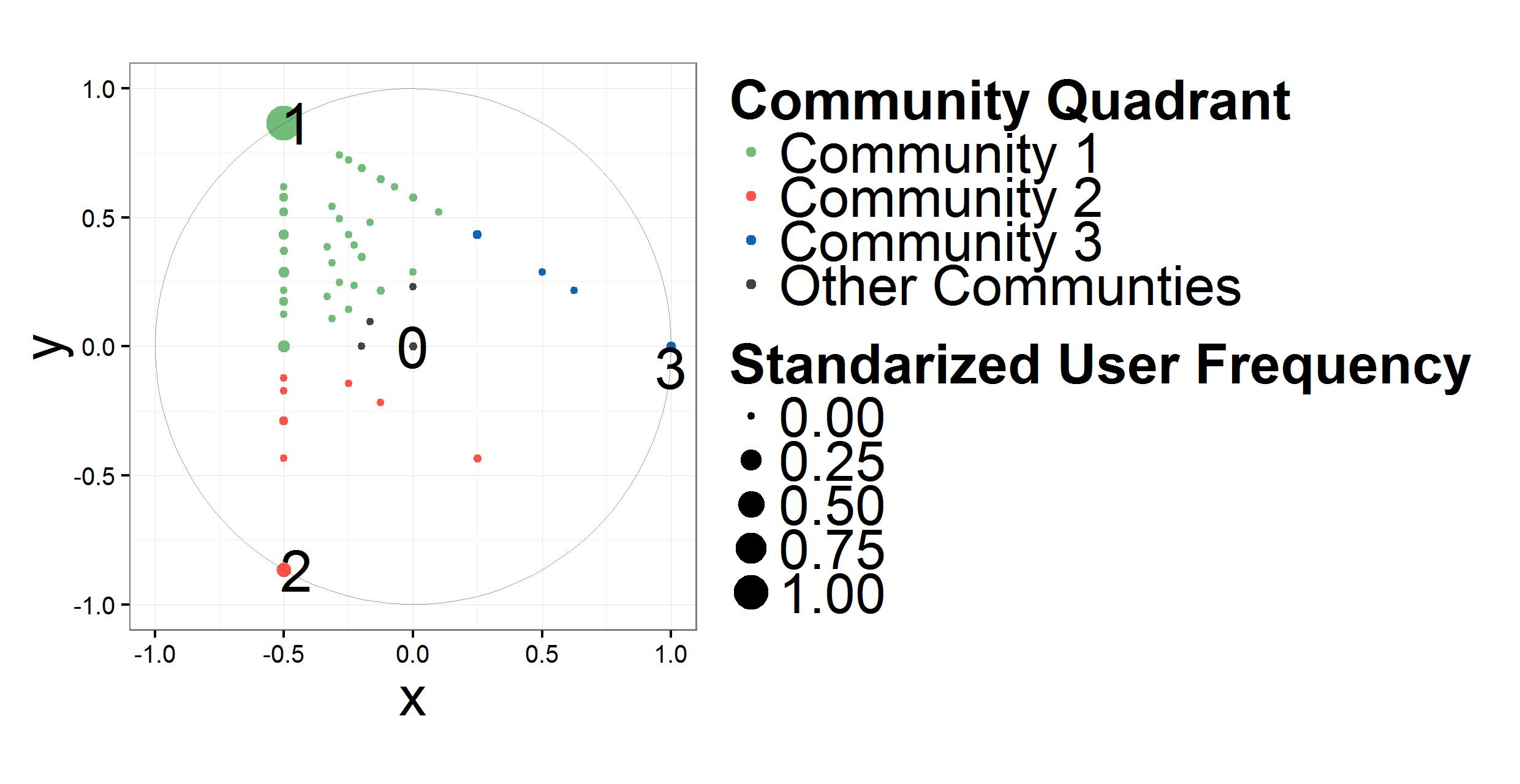}
		\caption{Twitter}
		\label{fig:twitter_polarization}
	\end{subfigure}
	\caption{Users Polarization: Users activity across the three most active communities of Facebook a) and Twitter b). In the graph, the vertices of the triangle represent the three most active	communities and the central point all the remaining ones. The position of each dot is determined by the number of communities the users interacts with. The size indicates the
		number of users in that position. Hence, data denotes a very characteristic behavior: users are strongly polarized and tend to focus their attention on a single community of pages.}
	\label{fig:polarization}
\end{figure*}

\subsection{Communities and Users Polarization}

Online social media facilitate the aggregation of individuals in communities of interest, also known as echo chambers \cite{quattrociocchi2016echo,zollo2015debunking}, especially when considering users interaction around conflicting and contrasting narratives \cite{del2016spreading}. In this manuscript we do not perform any categorization of contents a priori. On the contrary, we only account for the connections created by users activities, and then observe the resulting, emerging communities.

We compare the results of the different community detection algorithms on all three bipartite projections, for the case of both Facebook and Twitter. Our analysis focuses on the the bipartite networks built around likes and favorites/retweets, since a) these activities represent more explicit positive feedbacks than comments or replies; b) the average Rand Index is greater for all the projections in both social media w.r.t. the bipartite networks built on comments; and c) in Twitter we lose fewer news sources accounts by applying the projections on the favorite bipartite network w.r.t. the comment one.

To compare each projection of the like/favorite bipartite networks we compute the average Rand Index and NMI among the partitions returned by the algorithms. For the case of both Facebook and Twitter the \projname obtains a better degree of concordance among the algorithms w.r.t. the simple edge counting projection. For instance, in Twitter the \projname gets $0.73$ and $0.69$ as average Rand index and NMI, while the simple edge counting gets $0.49$ and $0.28$, indicating a low concordance among the algorithms. Jaccard similarity performs even better (Rand = $0.8$ and NMI=$0.7$), however we choose the \projname since it accounts for the users' activity frequency and how they allocate their interactions among the pages they like. Similar results hold for the Facebook bipartite network, i.e., the \projname obtains $0.68$ and $0.66$ as Rand index and NMI, respectively. Finally, having fixed the bipartite network and its projection, we evaluate the concordance of each algorithm with the other ones. In Facebook the Louvain algorithm obtains the best average accordance (average Rand=$0.76$ and NMI=$0.65$), while in Twitter the algorithms based on dynamical processes get the best scores. The latter result is due to a high intra-approaches concordance and low inter-approaches scores, so that the biggest class drives the final score. To maintain a common setting between Facebook and Twitter, in the remaining of the paper we will consider and compare the communities extracted by the Louvain algorithm from the likes/favorites bipartite networks projected by the \projname.

In Facebook we identify five communities of news pages, while in Twitter the communities are four. In Figure \ref{fig:commmunities} we show the structure of Facebook and Twitter networks where vertices are grouped according to their community membership, while a complete list of the pages/accounts and their membership is reported in Table \ref{tab:pages}. For the case of both Facebook and Twitter, pages are not equally distributed among the communities, but we observe a main community ($C1$) including about $40\%$ of vertices and two smaller communities ($C2$ and $C3$), each one formed by $20\%$ of accounts. Activities on pages/accounts are even more skewed; the main community $C1$ accounts for most of the activities ($80\%$) in both Facebook and Twitter.

In order to characterize the relationship between the observed spontaneous emerging community structure of Facebook pages/ Twitter accounts and users behavior, we quantify the fraction of the activity of any user in the three most active communities versus that in any other community.
Figure \ref{fig:polarization} shows the activity of users across the three most active communities emerging from either Facebook pages (Fig. \ref{fig:facebok_polarization}) or Twitter accounts (Fig. \ref{fig:twitter_polarization}).
In both cases, we find that users are strongly polarized and that their attention is confined to a single community of pages. Both dynamics are similar, however we observe a substantial differentiation in the activity volume, with a smaller and even more polarized activity for the Twitter case.

\subsection{News Presentation}
We observed that users activity is mainly restricted to one echo chamber and that users show a limited interaction with other news sources. We are now interested in learning the process that drives such a pattern, which is observed in both Facebook and Twitter. We do that by measuring the distance between the sentiment with which the news is presented to the audience and the sentiment expressed by the users w.r.t. the same topic. To perform the analysis we make use of IBM Watson\texttrademark~AlchemyLanguage service API \cite{alchemy}, which allows to extract semantic meta-data from posts content. Such a procedure applies machine learning and natural language processing techniques aimed to analyze text by automatically extracting relevant entities, their semantic relationships, as well as the emotional sentiment they express \cite{gangemi2013comparison}. 
In particular, we extract the sentiment and the main entities presented by each post of our datasets, whether it has a textual description or a link to an external document. Entities are represented by items such as persons, places, and organizations that are present in the input text. Entity extraction adds semantic knowledge to content to help understand the subject and context of the text that is being analyzed. 

\begin{figure*}[h!]
	\begin{subfigure}{0.9\columnwidth}
		\centering
		\includegraphics[width=\textwidth]{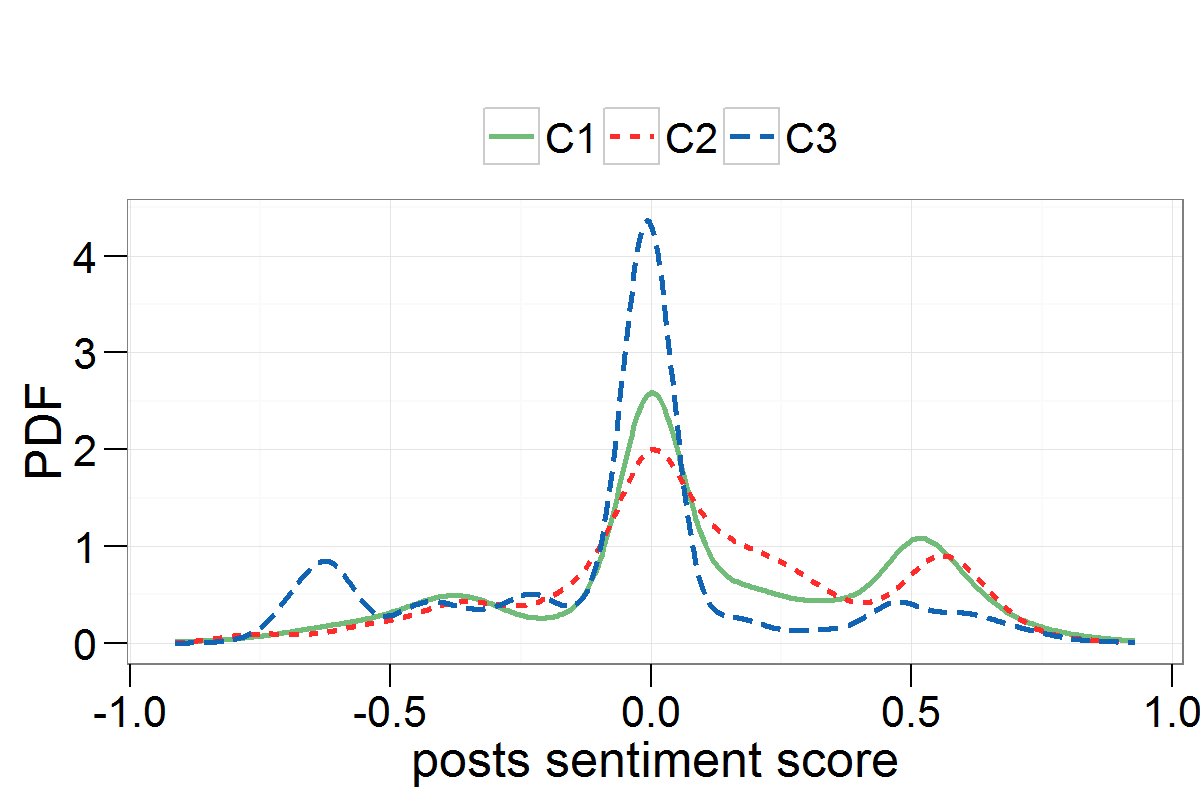}
		\caption{Facebook}
		\label{fig:facebok_sentiscore}
	\end{subfigure}
	\begin{subfigure}{0.9\columnwidth}
		\centering
		\includegraphics[width=\textwidth]{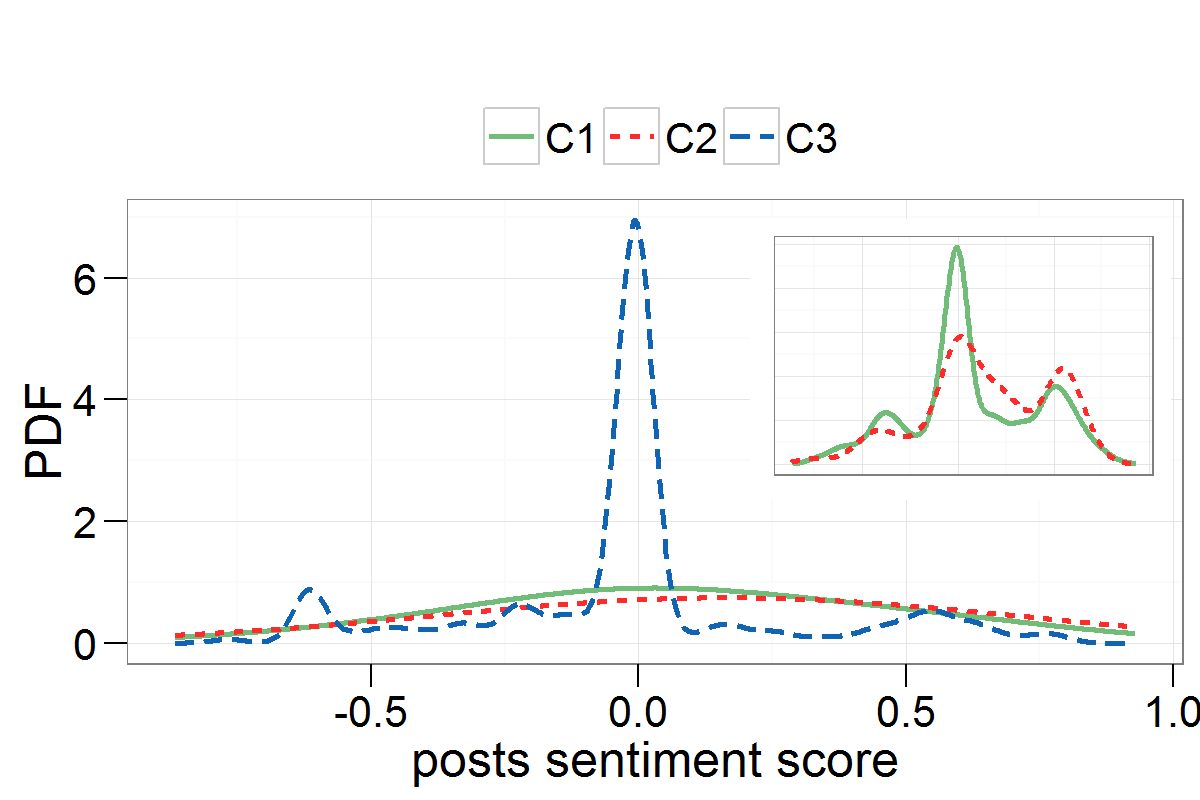}
		\caption{Twitter}
		\label{fig:twitter_sentiscore}
	\end{subfigure}
	\caption{Probability Density Function (PDF) of posts sentiment score on C1 (solid green), C2 (dashed red), and
		C3 (dashed blue) of Facebook a) and on Twitter b). The sentiment score is defined in the range $[-1, 1]$, where $-1$ is negative, $0$
		is neutral, and $1$ is positive.}
	\label{fig:sentiscore}
\end{figure*}
As a first step towards the identification and analysis of controversial topics, we look at how news are presented.
Figure~\ref{fig:sentiscore} shows the Probability Density Function of the posts sentiment score on the three most active communities for both Facebook (Figure \ref{fig:facebok_sentiscore}) and Twitter (Figure \ref{fig:twitter_sentiscore}). The sentiment score is defined in the range $[-1,1]$, where $-1$ is negative, $0$ is neutral, and $1$ is positive. A neutral overall pattern is observed on both platforms meaning that topics are mainly presented in a neutral manner, although exhibiting a slightly higher probability of positive sentiment on the two largest communities (solid green and dashed red lines) with respect to the smaller one. Such a behavior is observed on both Facebook and Twitter. Notice that we consider how subjects are presented in a post; here we do not take into account the sentiment that the post may elicit in the reader, or the sentiment of users involved in the discussion. 

\begin{figure*}[h!]
	\begin{subfigure}{1\columnwidth}
		\centering
		\includegraphics[width=\textwidth]{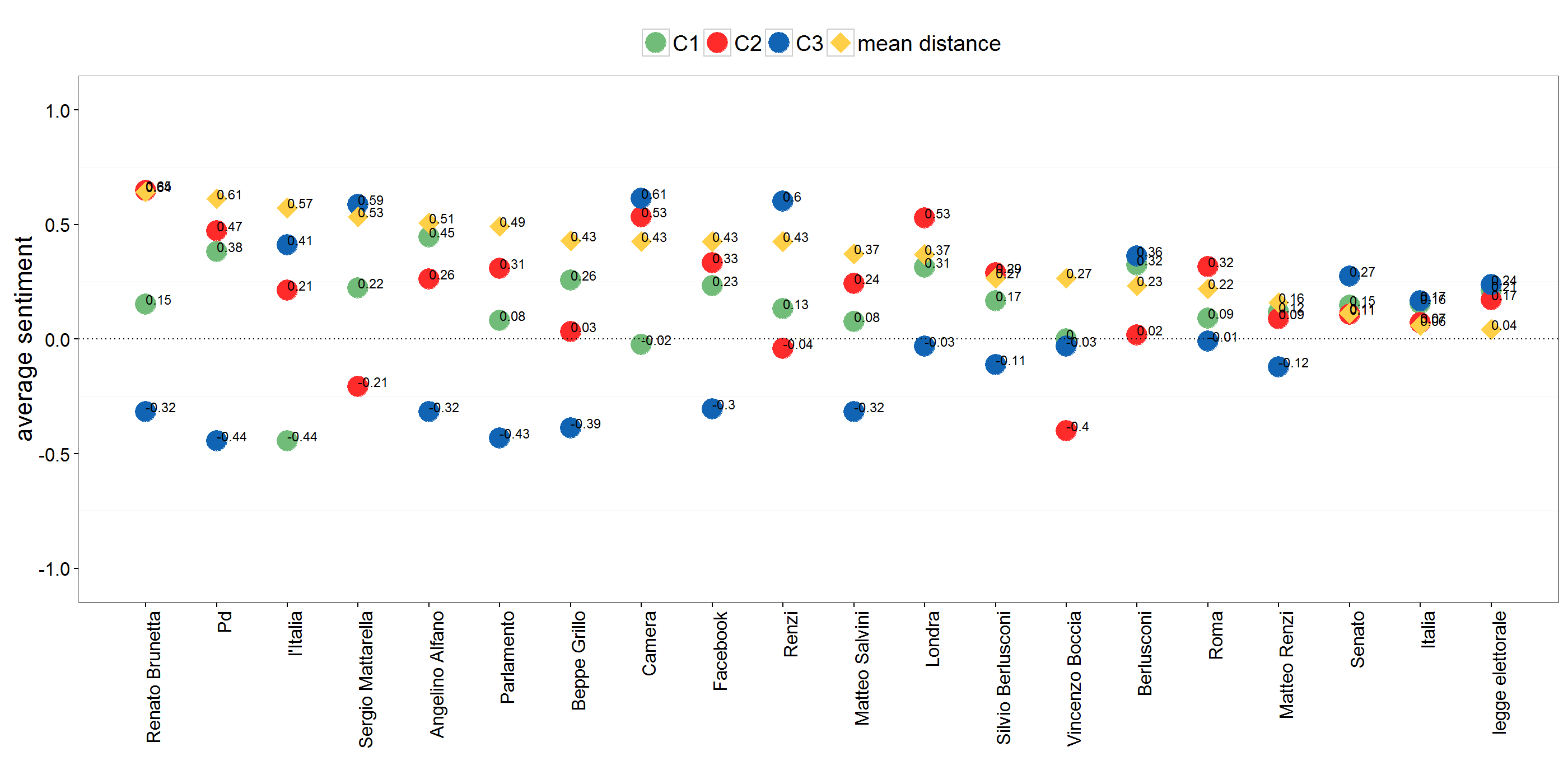}
		\caption{Facebook}
		\label{fig:emo_facebok_comp}
	\end{subfigure}
	\begin{subfigure}{1\columnwidth}
		\centering
		\includegraphics[width=\textwidth]{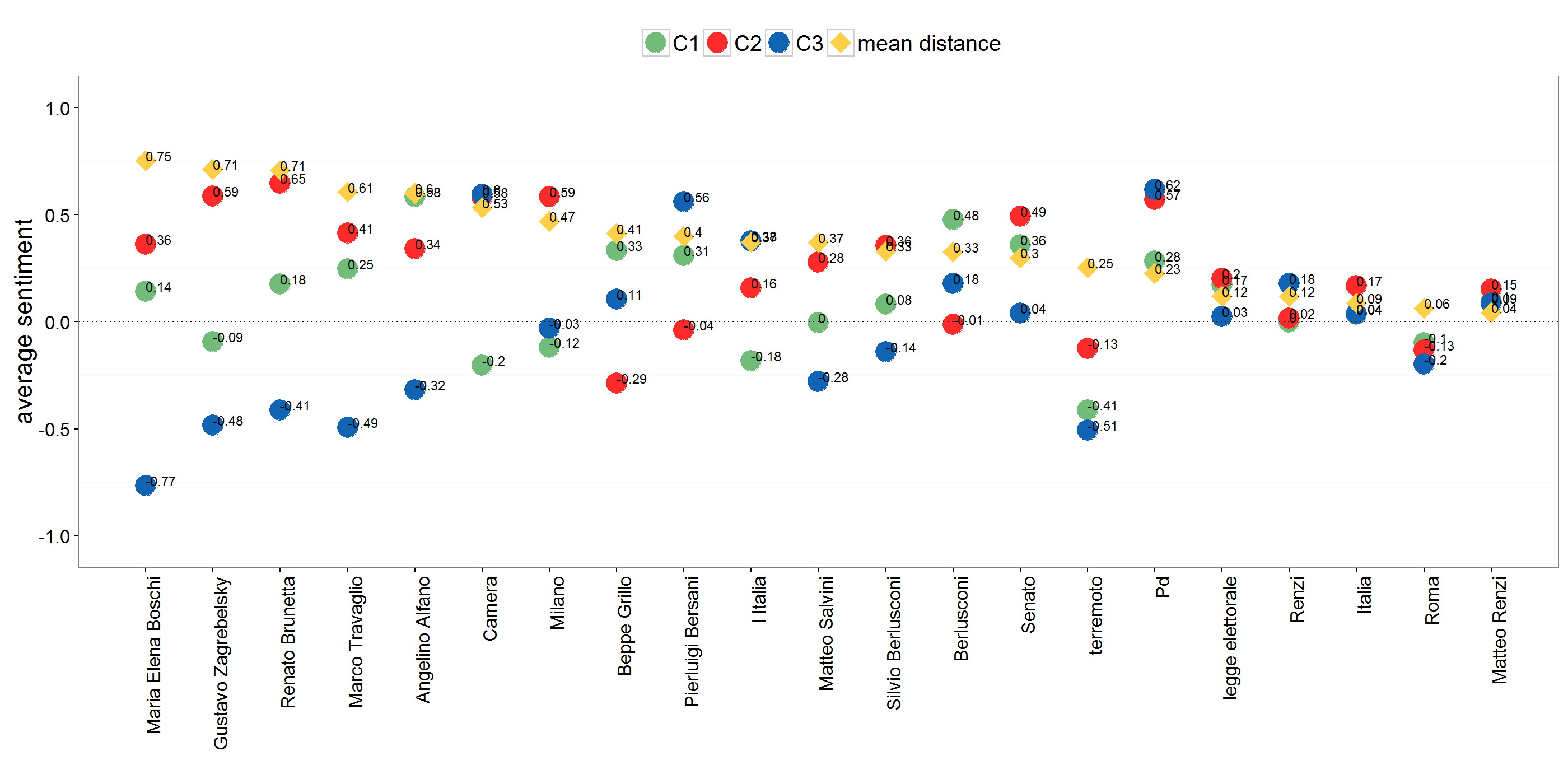}
		\caption{Twitter}
		\label{fig:emo_twitter_comp}
	\end{subfigure}	
	
	\begin{subfigure}{2\columnwidth}
		\centering
		\includegraphics[width=\textwidth]{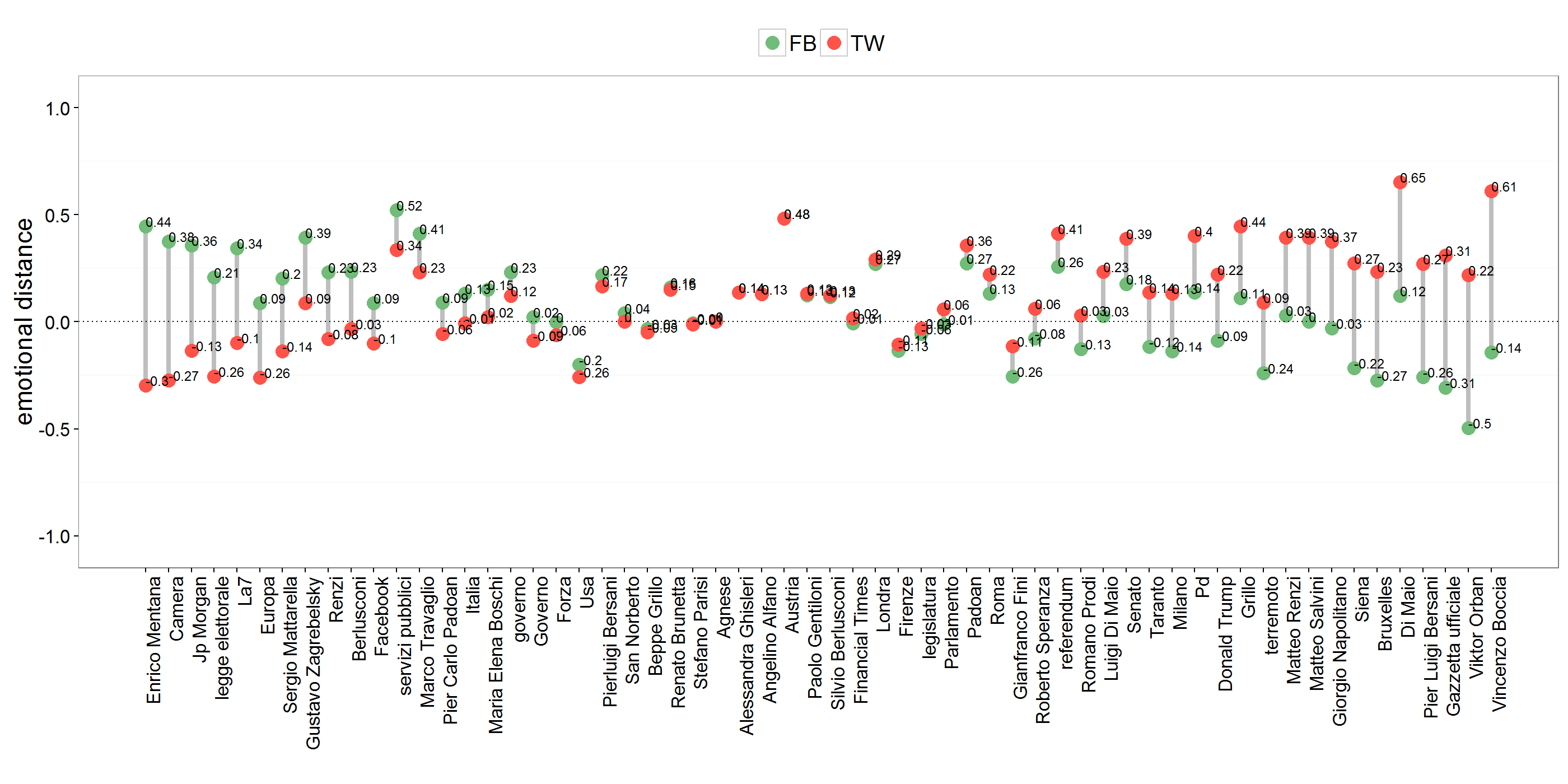}
		\caption{Facebook vs Twitter}
		\label{fig:entities_shared}
	\end{subfigure}		
	\caption{\textbf{Emotional Distance Among Communities.} Average sentiment of an entity on each community (green dots for C1, red for C2, and blue for C3) on Facebook a) and Twitter b).  Yellow diamonds indicates the mean emotional distance between pairs of communities, i.e., the mean of the differences between the average sentiment of an entity on each pair of communities, for each entity debated in all three communities. Entities are shown in a descending order by the largest to the smallest mean emotional distance. c) Emotional distance between Facebook communities aggregated (green dots) and Twitter one (red dots).}
	\label{fig:emotion_communities}	
\end{figure*}

We now focus on the entities shared across communities: on Facebook we identify $20$ such entities appearing in posts from all the three most active communities, while on Twitter they are $21$. For each entity we compute its average sentiment with respect to every community, i.e., we get three values representing the mean sentiment of the posts containing that entity in either $C1$, $C2$, or $C2$. The mean emotional distance of such entities is then the mean of the differences between the average sentiment pairwise computed. Figure \ref{fig:emo_facebok_comp} shows, for each entity, the average sentiment in each community (green dots fo $C1$, red for $C2$, and blue for $C3$) and the mean emotional distance (yellow diamonds) among communities on Facebook, while Figure \ref{fig:emo_twitter_comp} reports the same information for Twitter. In both panels a) and b) entities are shown in descending order with respect to their mean emotional distance, with those on the left being the ones discussed with the greatest difference in sentiment, while those on the right the ones discussed in a much more similar way across the three echo chambers. 

Figure \ref{fig:entities_shared} shows for each entity, the emotional distance between Facebook comunities aggregated (green dots) and Twitter ones (red dots). Entities on the left are presented in a more positive way on Facebook, and vice versa. We consider only entities appearing on both platforms and at least in two communities for each platform. 

\subsection{Users Response to Controversial Topics}
\begin{figure*}[h!]

	\begin{subfigure}{1\columnwidth}
		\centering
		\includegraphics[width=\textwidth]{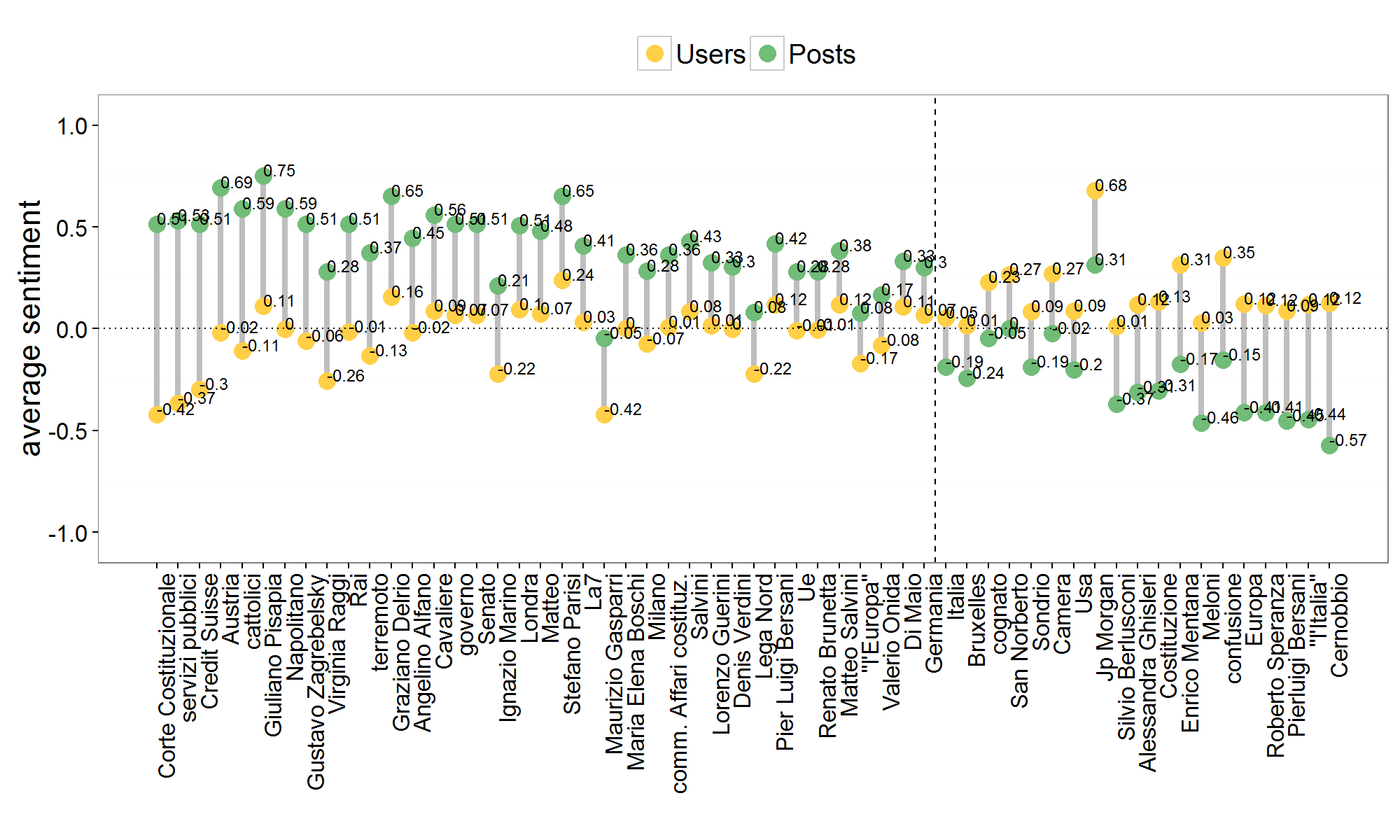}
		\caption{}
		\label{fig:emo_facebok_c1}
	\end{subfigure}
	\begin{subfigure}{1\columnwidth}
		\centering
		\includegraphics[width=\textwidth]{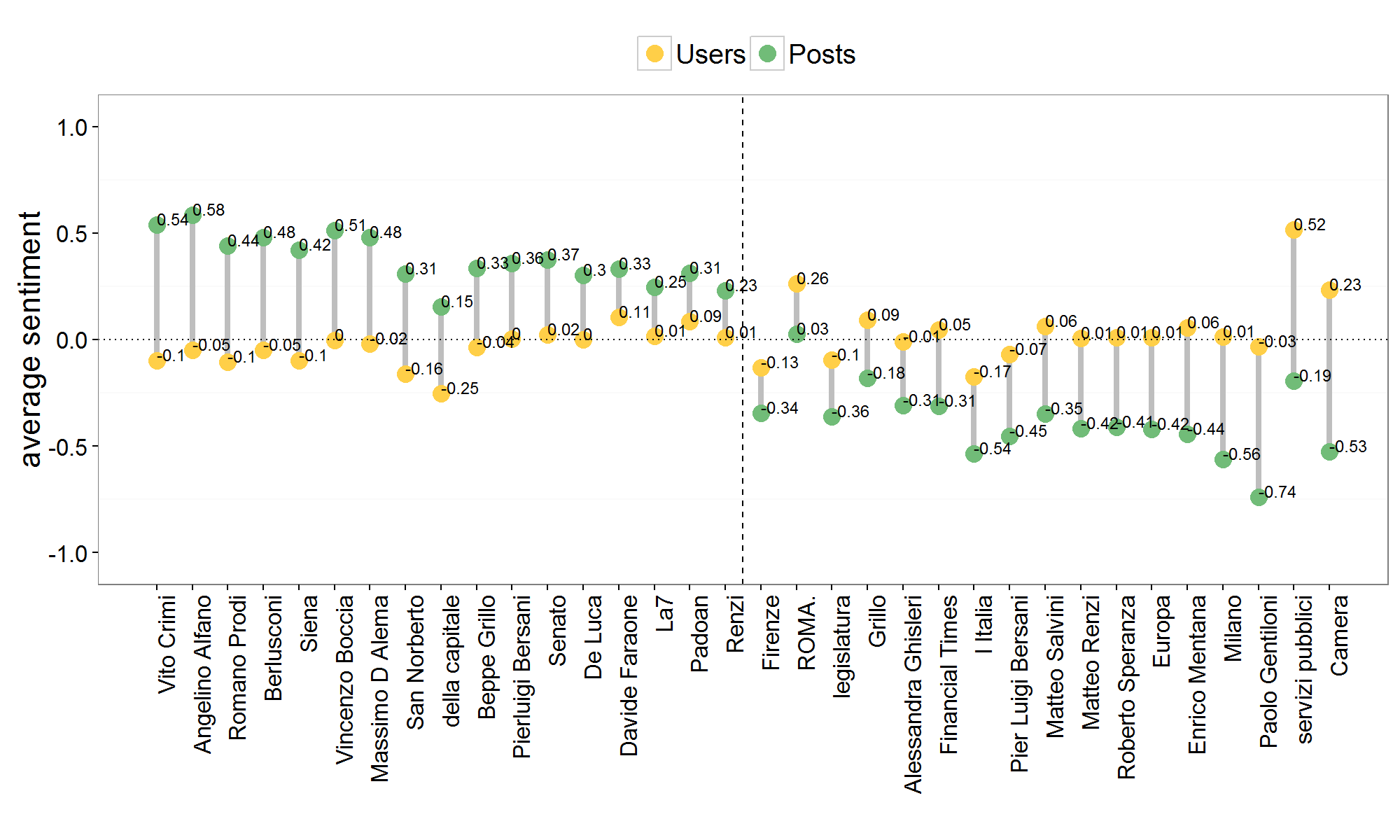}
		\caption{}
		\label{fig:emo_twitter_c1}
	\end{subfigure}
		
	\begin{subfigure}{1\columnwidth}
		\centering
		\includegraphics[width=\textwidth]{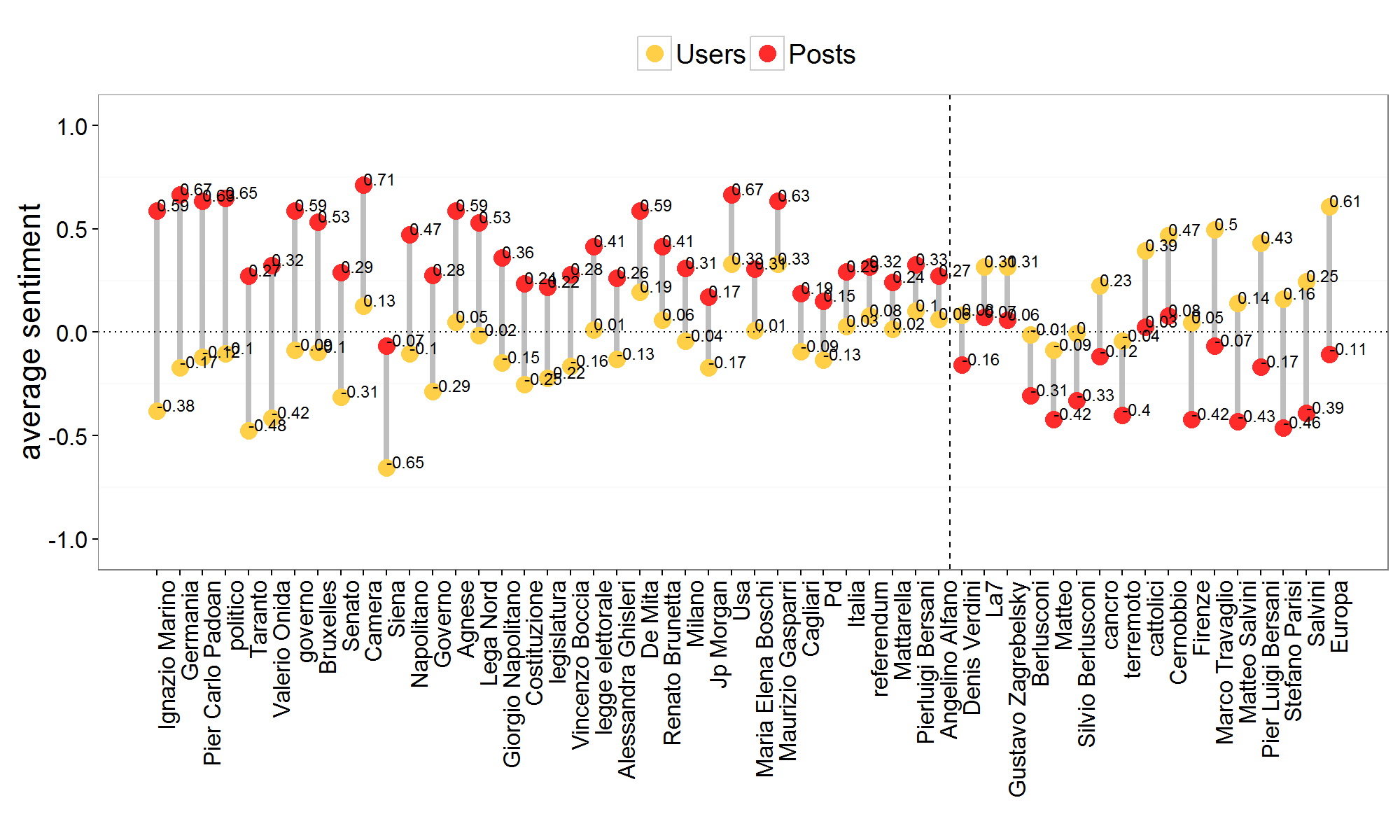}
		\caption{}
		\label{fig:emo_facebok_c2}
	\end{subfigure}	
	\begin{subfigure}{1\columnwidth}
		\centering
		\includegraphics[width=\textwidth]{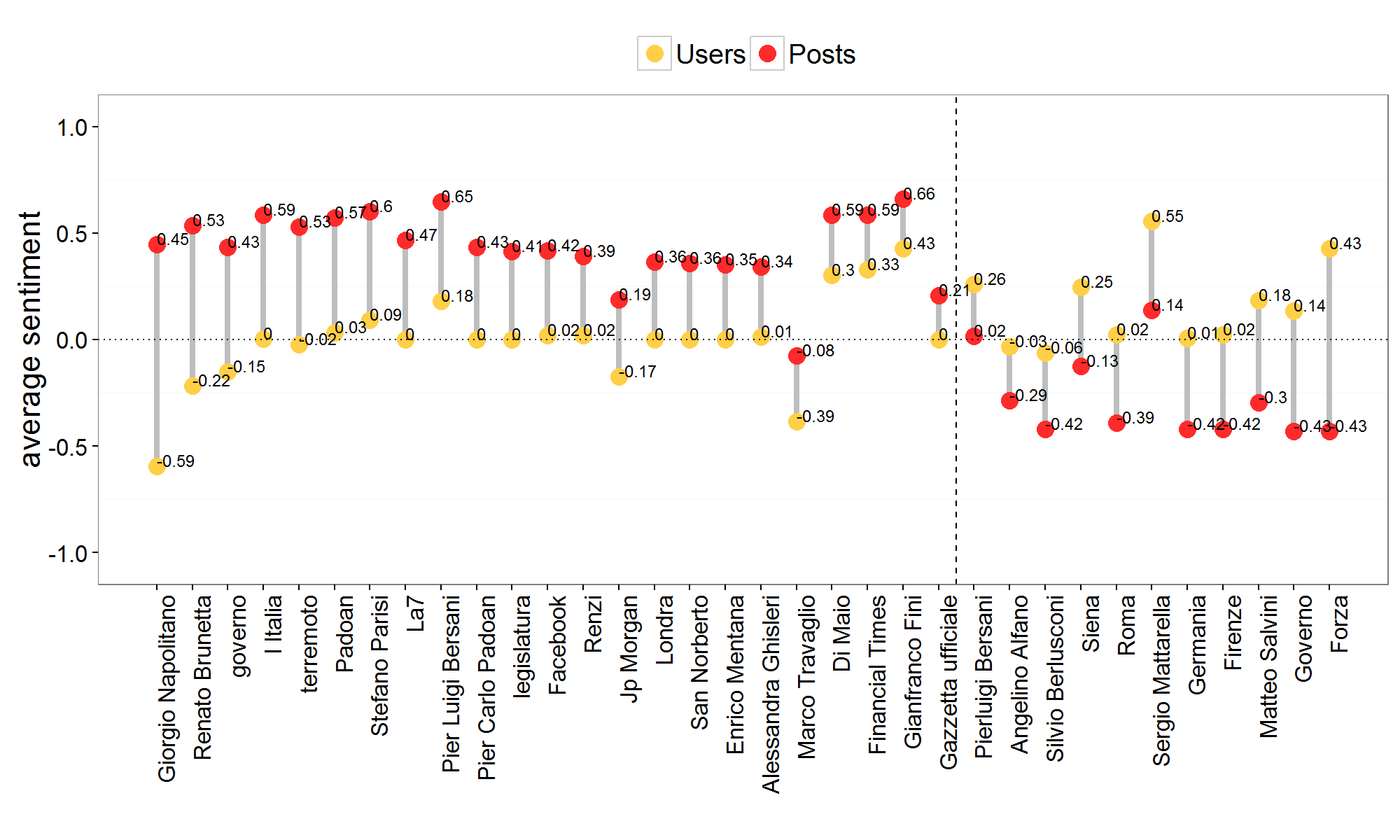}
		\caption{}
		\label{fig:emo_twitter_c2}
	\end{subfigure}
		
	\begin{subfigure}{1\columnwidth}
		\centering
		\includegraphics[width=\textwidth]{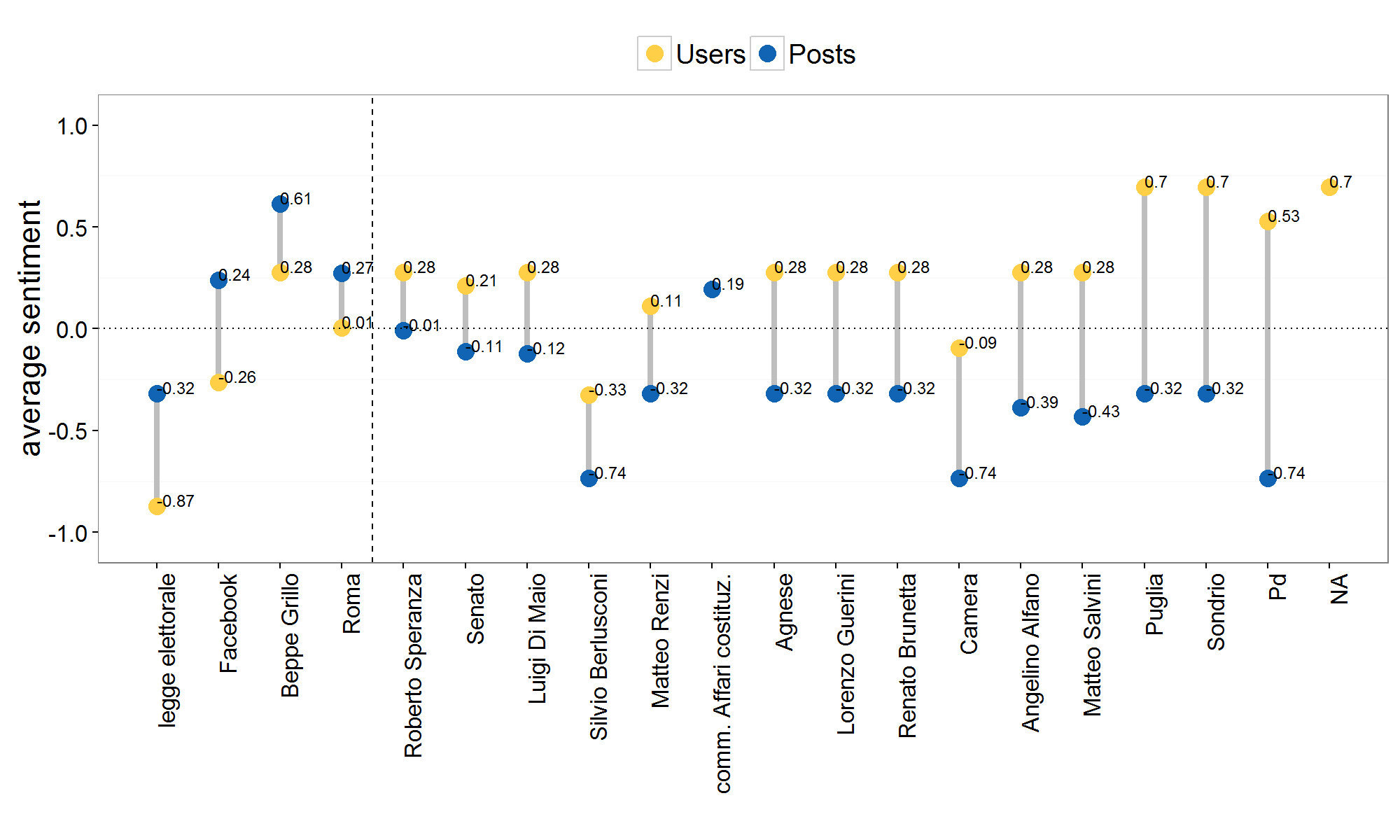}
		\caption{}
		\label{fig:emo_facebok_c3}
	\end{subfigure}
	\begin{subfigure}{1\columnwidth}
		\centering
		\includegraphics[width=\textwidth]{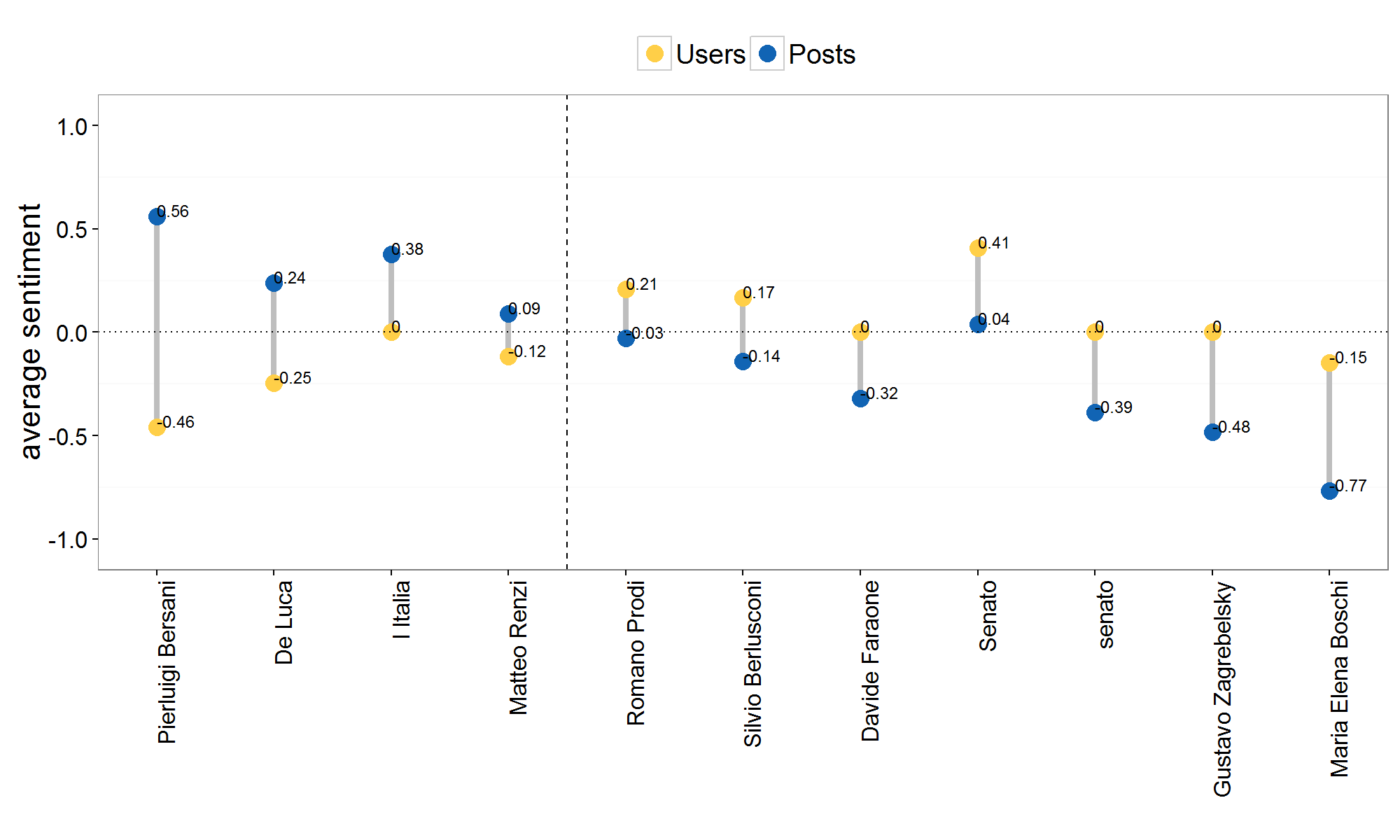}
		\caption{}
		\label{fig:emo_twitter_c3}
	\end{subfigure}
	\caption{Emotional Response to Controversial Entities. Left panels refer to Facebook, right ones to Twitter. In the specific, panels (a-b) show the emotional response of users (yellow dots) to posts of C1 (green dots) debating one of the listed controversial entities, panels (c-d) show the emotional response of users (yellow dots)
		to posts of C2 (red dots), and panels (e-f) show the emotional response of users (yellow dots) to posts of C3 (blue dots). Only entities for which the emotional distance between the posts and users sentiment is greater than $0.2$ are reported. The vertical dashed lines denote a change in users response.}
	\label{fig:emotion_comments}
\end{figure*}

We have characterized how subjects are debated inside and across communities and we identified the emerging controversial topics. We are now interested in the attitude of users towards such topics. We then analyze the emotional response of users by looking at the sentiment they express when commenting. Specifically, we focus on comments left on posts containing entities shared by at least two posts and we then compute their sentiment score through AlchemyAPI. Thus,to each comment is associated a sentiment score in $[-1, 1]$, characterized as before, and for each post (respectively, user) we compute the average sentiment of its (respectively, her) comments i.e., the mean of the sentiment of all comments of the post (respectively, user). Then, for each entity, we consider the emotional distance between the average sentiment of the post and that of its users. Since we are interested in identifying the most controversial entities, we consider only those for which the emotional distance (in absolute value) between the post and user is greater than $0.2$. Figure \ref{fig:emotion_comments} shows the emotional response of users to posts debating one of the listed controversial topics, on both Facebook and Twitter. Figures \ref{fig:emo_facebok_c1}, \ref{fig:emo_facebok_c2} and \ref{fig:emo_facebok_c3} (respectively, \ref{fig:emo_twitter_c1}, \ref{fig:emo_twitter_c2} and \ref{fig:emo_twitter_c3}) refer to $C1$, $C2$, and $C3$ in Facebook (respectively, Twitter). In all panels a vertical dashed line denotes a change in users response: entities on the left are those for which users response is more negative than the sentiment expressed in the post, and vice versa for those on the right. We may notice that users tend to react to the content of the posts in $C1$ and $C2$ slightly more negatively on Facebook rather than Twitter, while the opposite is observed for $C3$.

\textcolor{blue}{}

{\tiny
\begin{table}[ht]
	\centering
	\caption{\textbf{Facebook and Twitter News Sources with Community Membership}. News sources that are not members of any community are denoted by \textit{N} while those that are not available for Twitter are denoted by a dash.}
	\begin{tabular}{c |l | c| c} \hline
		\textbf{ID} & \textbf{Page Name} & \textbf{Facebook Community} & \textbf{Twitter Community}\\ \hline
		1 & Corriere della Sera & C1 & C1 \\
		2 & Il Fatto Quotidiano & C1 & C1 \\
		3 & Il Sole 24 Ore & C1 & C1 \\
		4 & La Repubblica & C1 & C1 \\
		5 & La Stampa & C1 & C1 \\
		6 & Corriere di Viterbo & C1 & C2 \\
		7 & La Nuova Ferrara & C1 & -\\
		8 & La Provincia di Como & C2 & C1 \\
		9 & Il Giornale & C2 & C2 \\
		10 & Libero Quotidiano & C2 & C2 \\
		11 & Quotidiano di Sicilia & C2 & C3 \\
		12 & Corriere di Siena & C2 & - \\
		13 & Corriere Adriatico & C3 & N \\
		14 & Il Centro & C3 & N \\
		15 & Il Piccolo & C3 & N \\
		16 & Il Tirreno & C3 & N \\
		17 & L'Eco di Bergamo & C3 & N \\
		18 & La Gazzetta di Parma & C3 & N \\
		19 & Il Messaggero & C3 & C1 \\
		20 & Corriere di Arezzo & C3 & C2 \\
		21 & Gazzetta di Modena & C3 & C2 \\
		22 & Il Resto del Carlino & C3 & C2 \\
		23 & Il Tempo & C3 & C2 \\
		24 & Giornale di Sicilia & C3 & C3\\
		25 & Il Mattino & C3 & C3 \\
		26 & Il Secolo XIX & C3 & C3 \\
		27 & La Nazione & C3 & C4 \\
		28 & Il Manifesto & C3 & - \\
		29 & La Sicilia & C3 & N \\
		30 & Il Messaggero Veneto& C4 & N \\
		31 & L'Unione Sarda& C4 & N \\
		32 & La Provincia di Cremona & C4 & N \\
		33 & Gazzetta di Reggio & C4 & C1 \\
		34 & La Gazzetta del Mezzogiorno & C4 & C1 \\
		35 & Giornale di Brescia & C4 & C3 \\
		36 & Italia Oggi & C4 & C3 \\
		37 & La Gazzetta di Mantova & C4 & C3 \\
		38 & Il Giorno & C4 & C4 \\
		39 & La Nuova Sardegna & C4 & C4 \\
		40 & Gazzetta del Sud & C4 & - \\
		41 & Il Giornale di Vicenza & C5 & N \\
		42 & Il Mattino di Padova & C5 & N \\
		43 & La Nuova di Venezia e Mestre & C5 & N \\
		44 & Trentino & C5 & N \\
		45 & Alto Adige & C5 & C1 \\
		46 & Il Gazzettino & C5 & C1 \\
		47 & L'Adige & C5 & C4 \\
		48 & Corriere delle Alpi & C5 & N \\
		49 & L'Arena & C5 & - \\
		50 & La Provincia Pavese & C5 & - \\
		51 & La Tribuna di Treviso & C5 & N \\
		52 & Corriere di Rieti e della Sabina & C6 & C1 \\
		53 & Corriere dell'Umbria & C6 & C2 \\
		54 & La Provincia di Lecco & N & N \\
		55 & La Provincia di Sondrio & N & N \\
		56 & Quotidiano di Puglia & N & N \\
		57 & Avvenire & N & - \\
		\hline
	\end{tabular}
	\label{tab:pages}
\end{table}}

\section{Conclusions}
Social media have radically changed the paradigm of news consumption and the rising attention to the spreading of fake news and unsubstantiated rumors on online social media led researchers to investigate different aspects of the phenomenon. In this paper we aim to understand the main determinants behind content consumption on social media by focusing on the online debate around the Italian Constitutional Referendum. Throughout a quantitative, cross-platform analysis on both Facebook public pages and Twitter accounts, we characterize the structural properties of the discussion, observing the emergence of well-separeted communities on both platforms. Such a segregation is completely spontaneous; indeed, we find that users tend to restrict their attention to a specific set of pages/accounts.  
Finally, taking advantage of automatic topic extraction and sentiment analysis techniques, we are able to identify the most controversial topics inside and across both platforms, and to measure the distance between how a certain topic is presented in the posts/tweets and the related emotional response of users. Our novel approach may provide interesting insights for a) the understanding of the evolution of the core narratives behind different echo chambers and b) the early detection of massive viral phenomena around false claims on online social media.

\bibliographystyle{ACM-Reference-Format}
\bibliography{biblio} 

\end{document}